\documentclass[12pt]{iopart}
\usepackage{epsfig}
\usepackage{graphicx}

\begin{document}
\title{\boldmath Flavor symmetry breaking of the nucleon sea
in the statistical approach}
\author[a]{Claude Bourrely$^1$, Jacques Soffer$^1$$^,$$^2$}
\address{$^1$ Aix Marseille Univ, Universit\'e de Toulon, CNRS, CPT, 
Marseille, France}
\address{$^2$Physics Department, Temple University,
1925 N, 12th Street, Philadelphia, PA 19122-1801, USA}
\eads{\mailto{claude.bourrely@cpt.univ-mrs.fr},
\mailto{jacques.soffer@gmail.com}}

\begin{abstract}
The flavor structure of the nucleon sea provides unique information to test the
statistical parton distributions approach, which imposes strong relations between
quark and antiquarks. These properties for unpolarized and helicity distributions
have been verified up to now by recent data. We will present here some new updated results 
which are a real challenge, also for forthcoming
accurate experimental results, mainly in the high Bjorken-$x$ region.
\end{abstract}
\maketitle

\section{\label{Intro} Introduction}
The structure of the nucleon sea is an important topic which has been the
subject of several
relevant review papers \cite{kumano,Chang}.
In spite of considerable progress made in our understanding, several aspects
remain to be
clarified, which is the goal of this paper.
The properties of the light quarks  {\it u} and {\it d}, which are the main
constituants of the
nucleon, will be revisited in the quantum statistical parton distributions
approach proposed more than one
decade ago. It is well known that the {\it u}-quark dominates over the 
{\it d}-quark,
but for antiquarks, $SU(2)$ symmetry was assumed for a long time, namely the
equality for the
corresponding antiquarks, ${\bar u} = {\bar d}$, leading to the Gottfried sum
rule \cite{gottfried}.
However the NMC Collaboration \cite{NMC} found that this sum rule is violated,
giving
a strong indication that ${\bar d} > {\bar u}$. In the statistical approach we
impose relations
between quarks and antiquarks and we treat simultaneously unpolarized
distributions
and helicity distributions, which strongly constrains the parameters, 
a unique situation in the literature.  As we will
see, this powerful tool
allows us to understand, not only the flavor symmetry breaking of the light
sea, but also the
Bjorken-$x$ behavior of all these distributions and to make challenging
specific predictions for forthcoming experimental results, in particular in the
high-$x$ region.

\section{\label{Forma} Formalism}
Let us now recall the main features of the statistical approach for building
up the parton distributions function (PDFs).
The fermion distributions are given by the sum of two terms,
a quasi Fermi-Dirac function and a helicity independent diffractive
contribution:
\begin{equation}
xq^h(x,Q^2_0)=
\frac{A_{q}X^h_{0q}x^{b_q}}{\exp [(x-X^h_{0q})/\bar{x}]+1}+
\frac{\tilde{A}_{q}x^{\tilde{b}_{q}}}{\exp(x/\bar{x})+1}~,
\label{eq1}
\end{equation}
\begin{equation}
x\bar{q}^h(x,Q^2_0)=
\frac{{\bar A_{q}}(X^{-h}_{0q})^{-1}x^{\bar{b}_ q}}{\exp
[(x+X^{-h}_{0q})/\bar{x}]+1}+
\frac{\tilde{A}_{q}x^{\tilde{b}_{q}}}{\exp(x/\bar{x})+1}~,
\label{eq2}
\end{equation}
at the input energy scale $Q_0^2=1 \mbox{GeV}^2$.
We note that the diffractive
term is absent in the quark helicity distribution $\Delta q$, in the quark
valence contribution $q - \bar q$ and in $u - d$ if one assumes $\tilde A_u = \tilde A_d$.\\
In Eqs.~(\ref{eq1},\ref{eq2}) the multiplicative factors $X^{h}_{0q}$ and
$(X^{-h}_{0q})^{-1}$ in
the numerators of the first terms of the $q$'s and $\bar{q}$'s
distributions, was justified in our attempt to generate the transverse momentum
dependence of the PDFs..
The parameter $\bar{x}$ plays the role of a {\it universal temperature}
and $X^{\pm}_{0q}$ are the two {\it thermodynamical potentials} of the quark
$q$, with helicity $h=\pm$. They represent the {\it fundamental parameters} of
the approach. Notice that following the chiral properties of QCD, we have
$X_{0q}^h = - X_{0\bar q}^{-h}$ in the exponentials.
For a given flavor $q$ the corresponding quark and antiquark distributions
involve the free parameters, $X^{\pm}_{0q}$, $A_q$, $\bar {A}_q$,
$\tilde {A}_q$, $b_q$, $\bar {b}_q$ and $\tilde {b}_q$, whose number is reduced
to $\it
seven$ by the valence sum rule, $\int (q(x) - \bar
{q}(x))dx = N_q$, where $N_q = 2, 1 ~~\mbox{for}~~ u, d$, respectively.

{}From a fit of unpolarized and polarized experimental data we have obtained
for the potentials the values \cite{Bourrely:2015kla}:
\begin{equation}
X_u^+ = 0.475\pm 0.001, \quad X_u^- = X _d^- = 0.307\pm 0.001,  
\quad X_d^+ = 0.244\pm 0.001.
\label{potval}
\end{equation}
It turns out that two potentials have identical numerical values,
so  for light quarks we have found  the following hierarchy between the
different potential components
\begin{equation}
X_u^+ >  X_u^- = X _d^- >  X_d^+ .
\label{potherar}
\end{equation}
We notice that quark helicity PDFs increases with the potential value, while
antiquarks
helicity PDFs increases when the potential decreases.

The above hierarchy implies  the following hierarchy on the 
quark helicity distributions for any $x,Q^2$,
\begin{equation}
 xu_+(x,Q^2) > xu_-(x,Q^2) = xd_-(x,Q^2) > xd_+(x,Q^2)
\label{ineq}
\end{equation}
and also the obvious hierarchy for the antiquarks, namely
\begin{equation}
x\bar d_- (x,Q^2) >  x\bar d_+ (x,Q^2) = x\bar u_+ (x,Q^2) >  x\bar u_- (x,Q^2)
{}.
\label{ineqbar}
\end{equation}
\section{\label{Resu} Results}
\begin{figure}[htp]    % fig 1
\begin{center}
\includegraphics[width=6.0cm]{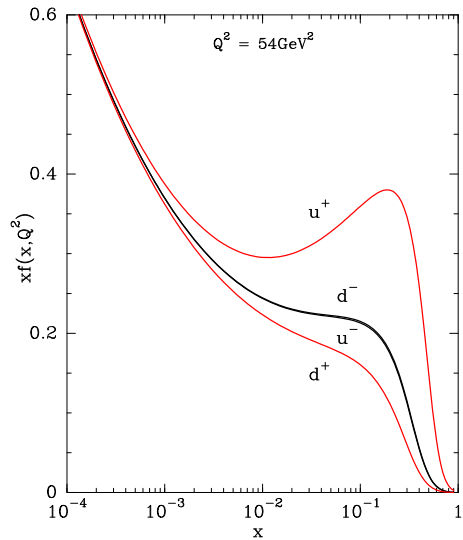}
\vspace*{+1.0ex}
\caption[*]{\baselineskip 1pt
 The different helicity components of the light quark distributions $xf(x,Q^2)$
($f=u_+, u_- = d_-, d_+$), versus $x$,
 at $Q^2= 54 \mbox{GeV}^2$, after NLO QCD evolution, from the initial scale
$Q_{0}^2 = 1 \mbox{GeV}^2$.}
\label{quarkhel}
\end{center}
\end{figure}

\begin{figure}[hpt]   % fig 2
\begin{center}
\includegraphics[width=6.0cm]{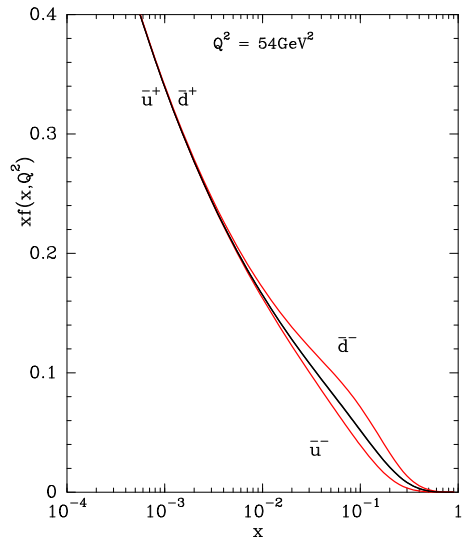}
\vspace*{+1.0ex}
\caption[*]{\baselineskip 1pt
The different helicity components of the light antiquark distributions
$xf(x,Q^2)$ ($f=\bar d_-, \bar d_+ = \bar u_+, \bar u_-$),
versus $x$, at $Q^2= 54 \mbox{GeV}^2$, after NLO QCD evolution, from the
initial scale $Q_{0}^2 = 1 \mbox{GeV}^2$.}
\label{antiquarkhel}
\end{center}
\end{figure}

\begin{figure}[htp]    % fig3
\begin{center}
\includegraphics[width=5.4cm]{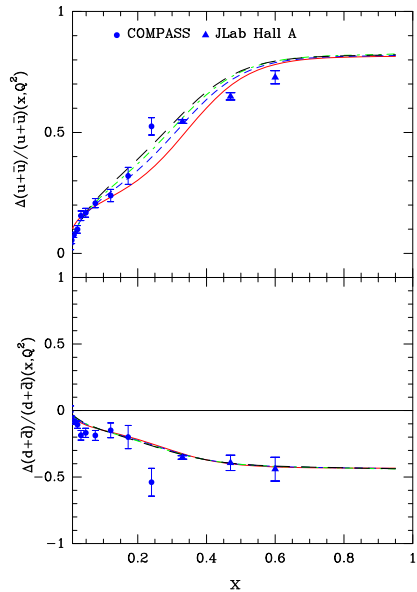}
\vspace*{+1.0ex}
\caption[*]{\baselineskip 1pt
Predicted ratios $(\Delta u(x,Q^2) + \Delta \bar u(x,Q^2)) /  (u(x,Q^2) + \bar
u(x,Q^2))$ and
$(\Delta d(x,Q^2) + \Delta \bar d(x,Q^2)) /  (d(x,Q^2 + \bar d(x,Q^2))$, versus
$x$, at $Q^2~(\mbox{GeV}^2) = 1 ~\mbox{solid}, 10 ~\mbox{dashed}, 100
 ~\mbox{dashed-dotted}, 1000 ~\mbox{long-dashed}$. Data are
from Refs. \cite{JLab} (Jlab) and \cite{compass} (COMPASS).}
\label{disratiosl}
\end{center}
\end{figure}

For illustration we show in Figs. \ref{quarkhel}-\ref{antiquarkhel},
 the resulting
distributions at  $Q^2=54 \mbox{GeV}^2$, a particular value which will be
explained later. 
It is important to note that these inequalities Eqs. (\ref{ineq},\ref{ineqbar})
are preserved by the next-to-leading order QCD evolution, at least 
outside the diffractive region, namely for $x>0.1$.
 We have also checked that the initial analytic form
Eqs.~(\ref{eq1},\ref{eq2}), is almost preserved by the $Q^2$ evolution with
some small changes of the parameters. This general pattern displayed in 
Figs. \ref{quarkhel}-\ref{antiquarkhel}, does not change much for different $Q^2$ values.
We also remark that the {\it largest} distribution is indeed $xu_{+}(x,Q^2)$, which has a distinct maximum
around $x=0.3$, a relevant feature as we will see below.
In our approach one can conclude that,
 $u(x,Q^2) > d(x,Q^2)$ implies a flavor symmetry breaking
of the light sea, i.e. $\bar d(x,Q^2) > \bar u(x,Q^2)$, which is clearly seen in Fig. \ref{antiquarkhel}.
 A simple interpretation
of this result is a consequence of the Pauli exclusion
principle,  based on the fact that the proton contains two $u$-quarks and only
one $d$-quark.
We now turn to
 more significant outcomes concerning the helicity
 distributions which follow from Eqs. (\ref{ineq},\ref{ineqbar}).
First for the $u$-quark
\begin{equation}
 x\Delta u(x,Q^2) > 0 \quad\quad  x\Delta \bar u(x,Q^2) > 0.
\end{equation}
Similarly for the $d$-quark
\begin{equation}
 x\Delta d(x,Q^2) < 0 \quad\quad  x\Delta \bar d(x,Q^2) < 0.
\end{equation}
So once more, quarks and antiquarks are strongly related since opposite signs
for the quark helicity distributions,
imply opposite signs for the antiquark helicity distributions, (see Eqs.~(\ref{eq1},\ref{eq2})),
at variance with the simplifying flavor symmetry assumption $x\Delta \bar
u(x,Q^2) = x\Delta \bar d(x,Q^2)$.\\
Our predicted signs and magnitudes have been confirmed  \cite{Bourrely:2015kla}
by the measured single-helicity
asymmetry $A_L$ in the $W^{\pm}$ production at BNL-RHIC from STAR
\cite{Adamczyk:2014xyw}. For the extraction
of helicity distributions, this process is expected to be clearner than
semi-inclusive DIS, because it does not involve
fragmentation functions.\\
Another important earlier prediction concerns the Deep Inelastic Scattering
(DIS) asymmetries, more precisely
$(\Delta q(x,Q^2) + \Delta \bar q(x,Q^2)) /  (q(x,Q^2) + \bar q(x,Q^2))$
($q=u,d$), shown in Fig. \ref{disratiosl}.
Note that the JLab \cite{JLab} data and the COMPASS  \cite{compass}  data 
are in agreement with these predictions, in
particular in the high-$x$ region where
there is a great accuracy.  Beyond $x=0.6$, this is a new challenge for the JLab
12 GeV upgrade, with an extremely high luminosity, will certainly reach a much better precision.
%\clearpage
\\
There are two more strong consequences of the equalities in Eqs.
(\ref{ineq},\ref{ineqbar}), which relate unpolarized
and helicity distributions, namely for quarks
\begin{equation}
xu(x,Q^2) -  xd(x,Q^2) = x\Delta u(x,Q^2) -  x\Delta d(x,Q^2)   > 0 ,
\label{relatqdq}
\end{equation}
and similarly for antiquarks
\begin{equation}
x\bar d(x,Q^2) - x\bar u(x,Q^2) = x\Delta \bar u(x,Q^2) -  x\Delta \bar
d(x,Q^2) > 0.
\label{relatqdqbar}
\end{equation}
These equalities mean
 that the flavor asymmetry of the light quark and antiquark distributions is
the same for the corresponding helicity
distributions, as noticed long time ago, by comparing the isovector contributions 
to the structure
functions $2xg_1^{(p-n)}(x,Q^2)$ and $F_2^{(p-n)}(x,Q^2)$, which are the differences on proton
and neutron targets \cite{bbs-rev}. 
\begin{figure}[htp]    % fig4
\vspace*{-0.5ex}
\begin{center}
\includegraphics[width=5.4cm]{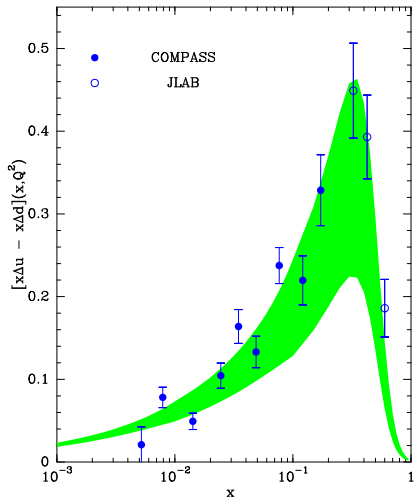}
\caption[*]{\baselineskip 1pt
The data on $x[\Delta u(x,Q^2) - \Delta d(x,Q^2)]$ 
from Refs. \cite{JLab} (JLab) and \cite{compass} (COMPASS), compared to the statistical model 
prediction, using Eq. (\ref{relatqdq}) with the corresponding error band. }
\label{diffq}
\end{center}
\end{figure}

\begin{figure}[htp]    % fig5
\vspace*{+0.5ex}
\begin{center}
\includegraphics[width=5.4cm]{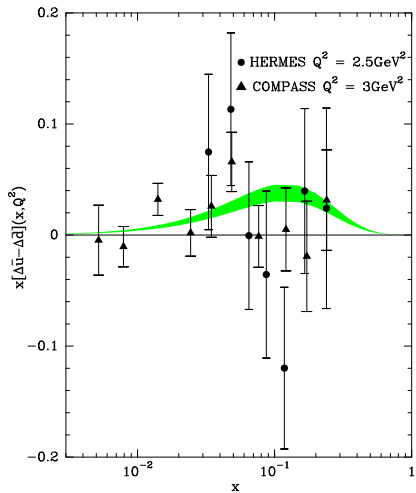}
\caption[*]{\baselineskip 1pt
The data on $x[\Delta \bar u(x,Q^2) - \Delta \bar d(x,Q^2)]$ 
from Refs. \cite{hermes} (HERMES) and \cite{compass} (COMPASS), compared to the statistical model 
prediction, using Eq. (\ref{relatqdqbar}) with the corresponding error band. }
\label{diffqbar}
\end{center}
\end{figure}

We have checked that using
Eq. (\ref{relatqdq}) it is possible to predict the helicity distributions from the unpolarized 
distributions, as displayed in Fig. \ref{diffq}. This difference, which is indeed positive, 
has a pronounced maximum
around $x=0.3$, reminiscent of the dominance of $xu_+(x,Q^2)$.\\
Similarly one can use Eq.  (\ref{relatqdqbar}), to predict the difference of the antiquark 
helicity distributions and the result is shown in Fig. \ref{diffqbar}.

\begin{figure}[htp]    % fig6
\begin{center}
\includegraphics[width=8.0cm]{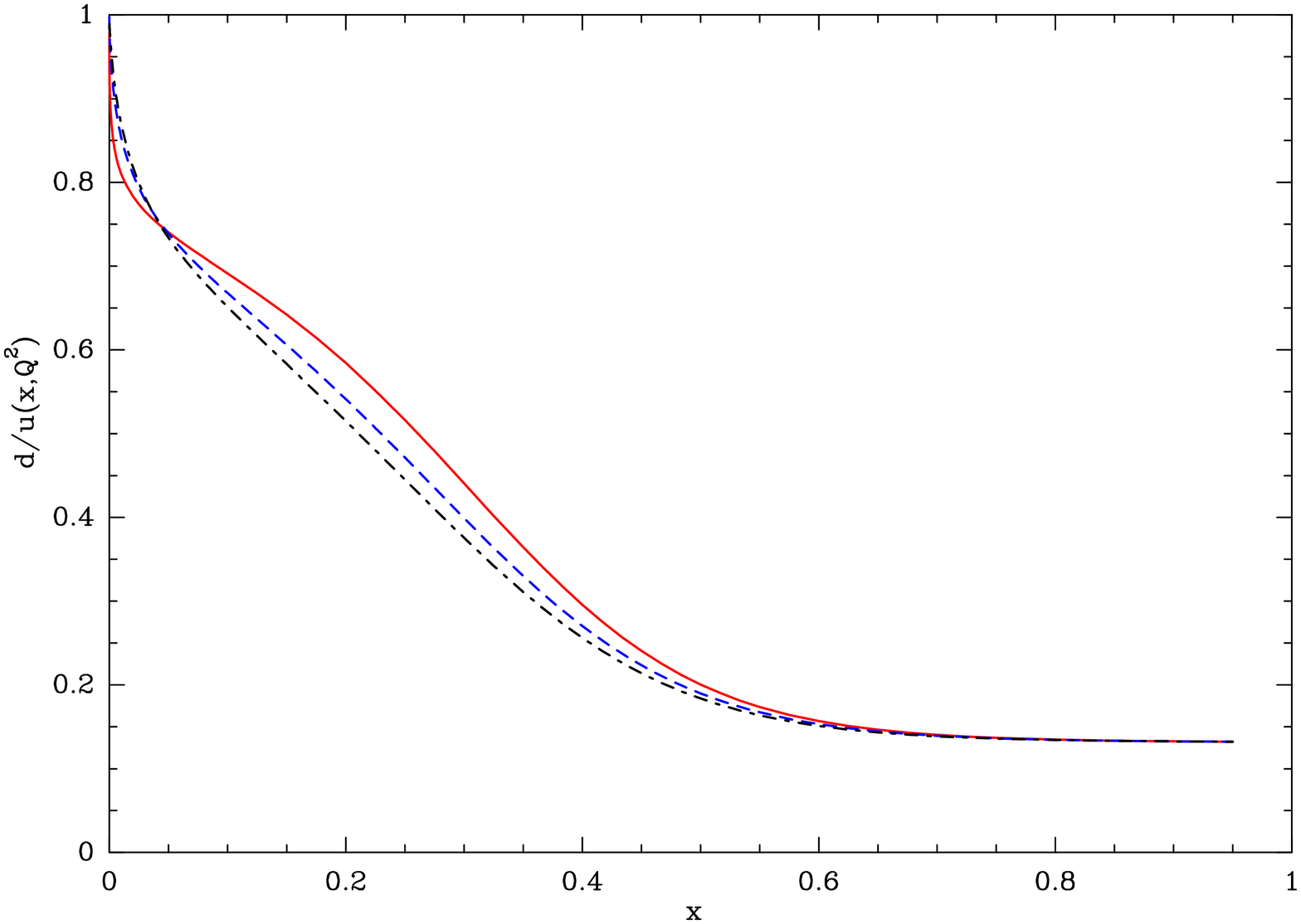}
\includegraphics[width=8.0cm]{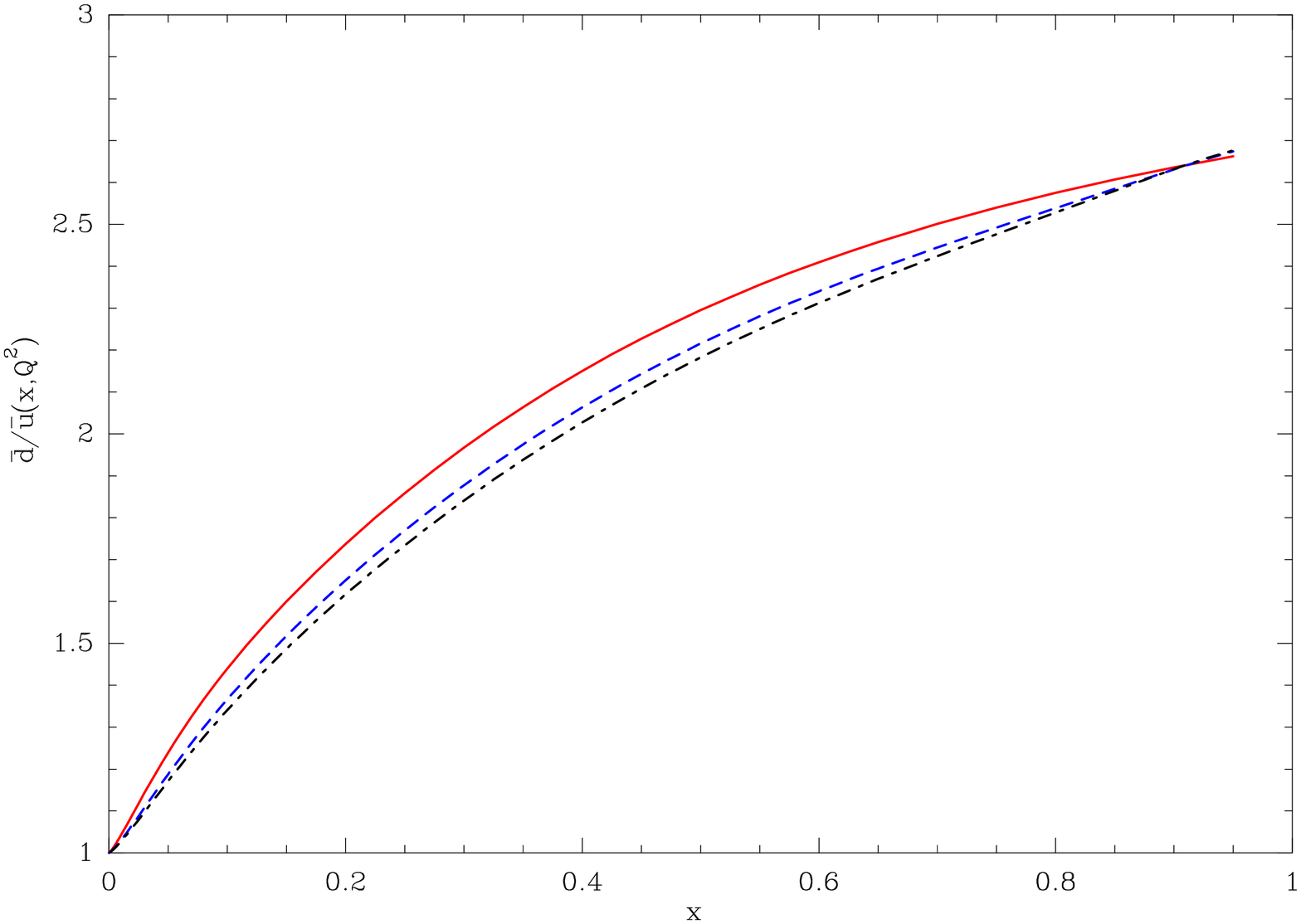}
\caption[*]{\baselineskip 1pt
The ratios $d(x,Q^2)/u(x,Q^2)$ (${\it top}$) and $\bar d(x,Q^2)/\bar u(x,Q^2)$
(${\it bottom}$) versus $x$ for 
$Q^2~(\mbox{GeV}^2)$ = 1~\mbox{solid}, 10 ~\mbox{dashed}, 100 ~\mbox{dashed-dotted} 
 }.
\label{ratios}
\end{center}
\end{figure}

\begin{figure}[hpb]  % fig 7 
\vspace*{-0.60ex}
\begin{center}
\includegraphics[width=6.2cm]{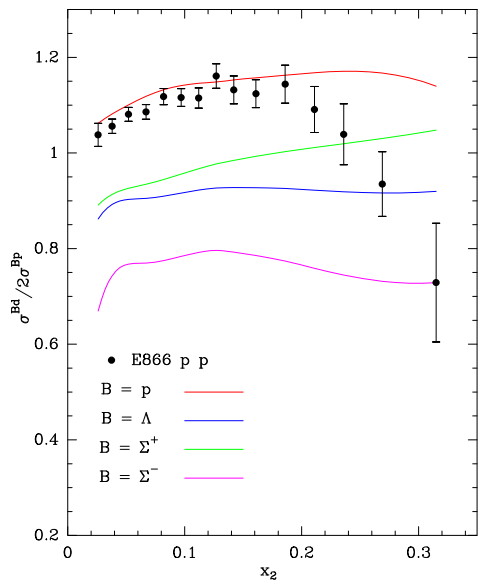}
\vspace*{+2.0ex}
\caption[*]{\baselineskip 1pt
The Drell-Yan cross section ratios of $\sigma(Bd)/2\sigma(Bp)$ versus $x_2$ (momentum fraction 
of the target partons) data from  (E866) Ref.  \cite{E866}. The curves are the statistical model predictions 
for different incoming beams $ B = p, \Lambda, \Sigma^+, \Sigma^-$
at $p_{lab} = 800$GeV.}
\label{rdy}
\end{center}
\end{figure}

\begin{figure}[htpb]  % fig 8 
\vspace*{-0.60ex}
\begin{center}
\includegraphics[width=7.0cm]{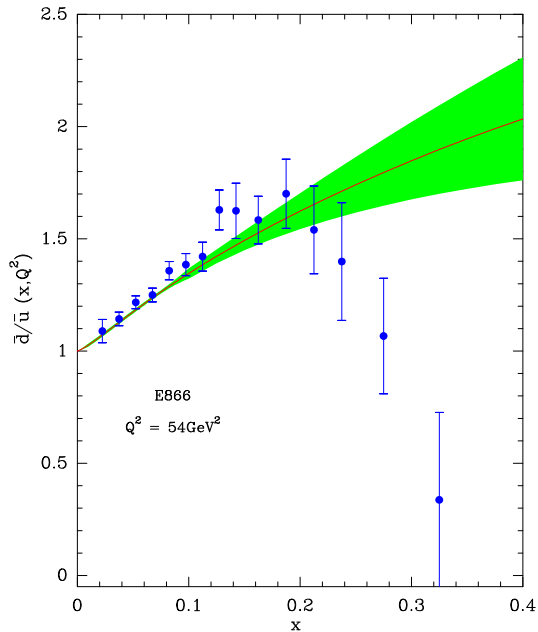}
\vspace*{+2.0ex}
\caption[*]{\baselineskip 1pt
The ratio $\bar d(x,Q^2)/\bar u(x,Q^2)$, versus $x$ for $Q^2= 54 \mbox{GeV}^2$.
The data from Ref.  \cite{E866} (E866)
are compared with the statistical model prediction with the corresponding error
band.}
\label{ratio}
\end{center}
\end{figure}

Although compatible with zero  it is slightly positive, but JLab 12GeV upgrade is expected to
reach a much better accuracy \cite{hafidi}.\\
Let us inspect the $x$-behavior of all these 
components $xu_+ (x, Q^2),..,x\bar u_-(x,Q^2)$, which are all
monotonic decreasing functions of $x$ at least for $x > 0.2$, outside the
region dominated by the diffractive contribution
(see Figs. \ref{quarkhel}-\ref{antiquarkhel}). As already said, $xu_+(x,Q^2)$
is the largest of the quark components and similarly $x\bar d_- (x\,Q^2)$ is the largest 
of the antiquark components.\\
The ratio $xd(x,Q^2)/xu(x,Q^2)$  value is one at $x=0$, because the
diffractive contribution
dominates and, due to the monotonic decreasing, it decreases for increasing
$x$. This $x$-behavior is strongly related to the values of the potentials
$X^h_{0q}$. \\
This falling $x$-behavior has been verified experimentaly from the ratio of the
DIS structure functions $F_{2}^{d} / F_{2}^p$
and from the charge asymmetry of the $W^{\pm}$ production in $\bar p p$
collisions \cite{Kuhlmann:1999sf}.\\
Similarly if one considers the ratio  $x\bar d(x,Q^2) /x\bar u(x,Q^2)$, its
value is one at $x=0$, because the diffractive
contribution dominates and, due to the
slightly larger value of $\bar d^-$ over $\bar u^-$, it increases for
increasing $x$ (see Fig. \ref{antiquarkhel}).\\
By looking at the curves (See Figure \ref{ratios}), one sees similar
behaviors. In both cases in the vicinity
of $x=0$ one has a sharp behavior due to the fact that the diffractive
contribution dominates. In the high-$x$ region
there is a flattening out above $x \simeq 0.6$ and it is remarkable to see that
these ratios have almost no $Q^2$ dependence.\\
In the introduction we have recalled the first indication by the NMC Collaboration
for a flavor asymmetry of the nucleon sea $\bar d(x) >\bar u(x)$. There is another 
way to probe this asymmetry, which is the ratio of the proton-induced Drell-Yan
process $\sigma(pd)/2\sigma(pp)$ on a deuterium and an hydrogen targets. At forward rapidity
region, the Drell-Yan cross section is dominated by the annihilation of a $u$-quark in the incident proton
with the $\bar u$-antiquark in the target. Assuming charge conjugaison one can show that
\begin{equation}
\sigma(pd)/2\sigma(pp) \sim 1/2[ 1 + \bar d(x_2)/\bar u(x_2)],
\label{dy}
\end{equation}
where $x_2$ refers to the momentum fraction of antiquarks.\\
The major advantage of the Drell-Yan process is that it allows to determine the 
$x$ dependence of $\bar d/\bar u$.
We show in Fig. \ref{rdy} the E866 from Ref.  \cite{E866} compared to our prediction. 
By assuming SU(3) symmetry
it is possible to generate the PDFs for the baryon octet and to calculate the 
corresponding Drell-Yan cross section
ratios. The results shown in Fig. \ref{rdy} might be of interest for future 
hyperon beams at LHC in fixed-target mode,
to study the sea structure of the hyperons. 
In the $\Lambda$
one expects no flavor symmetry breaking since it contains
one $u$-quark and one $d$-quark and under SU(3) symmetry one has,
$\bar u_{\Lambda} = \bar d_{\Lambda}= (\bar u +\bar d)/2$.
This is not the case for the $\Sigma^{\pm}$ and under SU(3) symmetry one expects
$\bar u_{\Sigma}/\bar s_{\Sigma} = \bar u/\bar d$, but this prediction
can be largely modified (see Ref. \cite{ahjt}).\\ 
The extracted ratio $\bar d(x)/\bar u(x)$ for $Q^2= 54 \mbox{GeV}^2$
displayed in Fig. \ref{ratio}, shows a remarkable agreement with the statistical model prediction 
up to $x=0.2$.

\newpage
\section{\label{Conclu} Conclusions}
To conclude a monotonic increase of the ratio $x\bar d (x,Q^2) /x\bar u(x,Q^2)$
is predicted \footnote{ It is interesting to recall that this
trend was already detected in a primitive version of the statistical model
\cite{bs95} (see Fig. 11).} in our approach, as a
consequence of strong relations between polarized quark distributions (see Eq.
(\ref{relatqdqbar})) . 
Very recently there was a serious indication from the preliminary results of
the SeaQuest collaboration \cite{reimer},
that this ratio rises beyond $x=0.2$, at variance with several
other model predictions,
as reported in Figs. 7 and 8 of Ref. \cite{Chang}. \\
This prediction is a real challenge for the statistical approach, whose strong
predictive power will be
confronted with several other forthcoming accurate data, mainly in the high-$x$
region, a region which remains poorly known.
\ack{
\noindent
J.S is very grateful to Prof. Norbert Vey and his team (IPC Marseille) for making possible the completion of this work.}

%%%%%%%%%%%%%%%%

\end{document}